# Graph Network Modeling Techniques for Visualizing Human Mobility Patterns

Sinjini Mitra
smitra16@asu.edu
Geometric Media Lab, Arizona State University
Tempe, United States

Anuj Srivastava
anuj@stat.fsu.edu
Department of Statistics, Florida State University
Tallahassee, United States

Avipsa Roy
avipsar@uci.edu
STaNCe Lab, University of California
Irvine, United States

Pavan Turaga
pturaga@asu.edu
Geometric Media Lab, Arizona State University
Tempe, United States

## Abstract

Human mobility analysis at urban-scale requires models to represent the complex nature of human movements, which in turn are affected by accessibility to nearby points of interest, underlying socioeconomic factors of a place, and local transport choices for people living in a geographic region. In this work, we represent human mobility and the associated flow of movements as a graph. Graph-based approaches for mobility analysis are still in their early stages of adoption and are actively being researched. The challenges of graph-based mobility analysis are multifaceted – the lack of sufficiently high-quality data to represent flows at high spatial and temporal resolution whereas, limited computational resources to translate large volumes of mobility data into a network structure, and scaling issues inherent in graph models etc. The current study develops a methodology by embedding graphs into a continuous space, which alleviates issues related to fast graph matching, graph time-series modeling, and visualization of mobility dynamics. Through experiments, we demonstrate how mobility data collected from taxicab trajectories could be transformed into network structures and patterns of mobility flow changes, and can be used for downstream tasks reporting $\approx$ 40% decrease in error on average in matched graphs compared to unmatched ones.

## CCS Concepts

• **Information systems** → **Geographic information systems**; • **Mathematics of computing** → **Graph algorithms**.

## Keywords

urban mobility, mobility network analysis, big data, graph matching

## 1 Introduction

Human mobility analysis at city-scale is a problem of increasing interest in geospatial computational sciences. Researchers have been developing methods to address issues of change monitoring, similarity detection, and pattern recognition for several decades [3, 23, 22]. Just by tracing trajectory data, it has been established as a fact that changes in mobility patterns arise as manifestations of underlying socioeconomic processes at play, such as variations in income levels, infrastructural access, occurrences of natural hazards, and pandemic outbreaks. Such processes inherently give rise to high-dimensional representations and are spatially heterogeneous, which makes them difficult to detect and quantify using traditional modeling and analytical tools. In growing urban cities, capturing these dynamic underlying determinants of mobility patterns is especiallyy challenging for urban planning and development [5] [17]. This issue can be partly subverted for a city in its nascent stages of growth by having a benchmark to compare to (an existing urban city). However, this presents its own unique set of challenges as city scale and growth is unique and cannot always be translated and compared 1:1 across time.

Thus, there is a need for a scale-invariant, robust, and mathematically well-grounded way to represent and predict human mobility patterns. With recent advances in graph-based network analysis [10, 20, 16], modeling human mobility patterns at such large spatial and temporal scales has become a possibility. Graphs and networks are naturally effective in handling scale effects in complex social phenomena. However, the problem of properly re-contextualizing existing urban growth data for cities in the nascent stages of development still remains unanswered. In this paper, we repose the problem of modeling human mobility by (1) introducing tools from geometric analysis of graphs to develop methods for representation, (2) and prediction of human mobility patterns through a downstream task. As demonstrated in Figure 1, we consider two cities at different scales of urban growth and create their respective mobility networks in a continuous graph space ($\mathcal{G}$). The two graphs are then matched by a graph matching algorithm and is followed by a downstream task (link prediction using neural networks, in this case). Our approach provides a way to develop further statistical tests for a variety of practical problems for human mobility in growing cities including: (i) tracking human mobility (at multiple geo-spatial scales), (ii) matching mobility patterns across the temporal axis, and (iii) predictive modeling of mobility.

## 2 Proposed Method

There are two facets to studying human dynamics data: (1) the shapes of and structural connections between transportation networks (made up of streets, roads, and highways) that supports human movements, and (2) the hourly, daily movements of people relative to these structural constraints. When one compares human dynamics across cities and time periods, one has to take both these facets – transportation networks and human mobility – into account. In order to analyze such "big" data (transportation networks and human movements) jointly, we need to represent them as **annotated graphs**. The shape and connectivity of a transportation network form a graph $G$ and the mobility data provide time-varying functional annotations at nodes and edges (such as pickup/drop off locations, total number of trips and passengers between points of interest etc.) of these graphs $f$.

### 2.1 Graph Space Network Structures

To compare and analyze shape and connectivity of the transportation network $G$, requires a formal definition of a graph space $\mathcal{G}$, specification of statistical models on $\mathcal{G}$, and inferential theory for low- and high-sample size statistics. Let $\{G_\gamma = (E_\gamma, V_\gamma) | \gamma \in \Gamma\}$ denote a multi-resolution graph with $n_\gamma$ nodes or vertices $V_\gamma$ and pairwise edges $E_\gamma(i, j)$ at a resolution $\gamma \in \Gamma$. Graphs are often represented by their adjacency or Laplacian matrices, for quantification and statistical analysis. One then imposes a metric structure on these representations to compare and quantify structural differences between graphs, and to develop statistical models. An important issue in graph-theoretic approaches is that the ordering of spatial nodes is arbitrary and one introduces the action of the permutation group to perform node registration across graphs [11, 12, 8].

If $\mathcal{A}$ is the set of graph representations, say all adjacency matrices, and $\mathcal{P}$ is the permutation group acting on $\mathcal{A}$, then $\mathcal{G} \equiv \mathcal{A}/\mathcal{P}$ forms the quotient space of $\mathcal{A}$ modulo $\mathcal{P}$. Any metric on $\mathcal{A}$, that is invariant to the action of $\mathcal{P}$, descends to $\mathcal{G}$ and provides a metric structure for quantitative analysis. In this setup, the node attributes can be geographical coordinates or transportation variables, such as pickup/drop off locations etc., while edge attributes can be anything from simple binary (connected or not connected) to the curvilinear shapes of road networks. Previous studies have used trip data on taxicabs to assess the vulnerability of movement patterns across different socioeconomic groups based on a network trip level geographically weighted regression of trip duration in the context of hurricane Sandy [18]. However, the study did not use graph matching or shape analysis for statistical modeling.

A recent work by Guo *et. al.* [8] uses this quotient structure to compute summary statistics, perform PCA-based dimension reduction, and to impose formal statistical models on $\mathcal{G}$. This framework naturally incorporates graph registration [4, 21, 2, 7, 15] in the analysis so that node-to-node registration can be inferred automatically. In this paper, we apply the methods of Guo *et. al.* [8] to the New York City Cab dataset, and discuss the affordances and interpretations of using Riemannian graph-metrics to measure change in transportation networks. In this setup, each graph is represented by the pair $(A, v)$ where $A$ is its adjacency matrix and $v$ is the vector of node attributes. The ordering of nodes in $(A, v)$ is arbitrary but compatible within the pair.

### 2.2 Graph Matching

Let $(A_1, v_1)$ and $(A_2, v_2)$ represent two graphs $G_1, G_2$ representing two different cities across time (for instance, Phoenix, AZ in 2017 and NYC in 2002). First assume that the two graphs have the same number of nodes, $n$. Let $\mathcal{P}$ be the set of all $n \times n$ permutation matrices. The problem of registering nodes across the two graphs can then be defined in terms of finding the best permutation of nodes in one graph that map to the nodes in the other graph. Formally, this is defined as finding the minima of the following cost function, over the set of all permutation matrices:

$$\hat{P} = \arg\min_{P \in \mathcal{P}} \left( \lambda \|A_1 - PA_2P^T\| + (1-\lambda)\text{Tr}(PD) \right), \quad (1)$$

where, $D$ is the matrix of all pairwise distances between node attributes across the two graphs, and $\lambda > 0$ is a scalar that balances the contributions of nodes and edges in matching graphs. Equation (1) represents the classic graph-matching problem and the literature provides a number of efficient yet approximate solutions (c.f. [21]). In this paper, we use the Fast Approximate Quadratic Programming (FAQ) method [21] with a gradient search to approximate the solution to Equation (1). Note that the minimum obtained in Equation (1) is a proper metric on the graph space $\mathcal{G}$; we denote it as $d_g(G_1, G_2)$. The FAQ algorithm restates the matching problem according to:

$$\min_{P \in \mathcal{P}} \left\| PA_1P^T - A_2 \right\|^2 = \min_{P \in \mathcal{P}} \left( -\text{Tr}\left(A_2PA_1P^T\right)\right) \quad (2)$$

For graphs with different number of nodes, we append the graphs with *null* nodes to bring them to the same dimensions. Null nodes are fictitious nodes that are assigned variable attributes to help optimize matching. Let $n_1$ and $n_2$ be the number of nodes in $G_1$ and $G_2$. We add $n_2$ and $n_1$ *null nodes* to $G_1$ and $G_2$ to bring the number of nodes to $n_1 + n_2$ in both graphs. We append the elements $(A_i, v_i)$ with zeros to reach the larger pair $(\tilde{A}_i, \tilde{v}_i)$ and then apply Equation (1) to match the appended graphs. A matching of a real node in the first graph to a null node in the second represents killing (or birth going the other way) of real nodes when going from first to the second.

To evaluate matching performance, we calculate the similarity between $G_1$ and $G_2$, by quantifying the difference between the two adjacency matrices $A_1, A_2$. For any two $A_1, A_2 \in \mathcal{A}$, with corresponding elements $a_{ij}^1$ and $a_{ij}^2$ respectively,

$$d_a(A_1, A_2) \equiv \sqrt{\sum_{i,j} d_m\left(a_{ij}^1, a_{ij}^2\right)^2} \quad (3)$$

where $d_a(A_1, A_2)$ quantifies the difference between $G_1$ and $G_2$. Here $d_m$ is the Riemannian distance on $\mathcal{M}$.

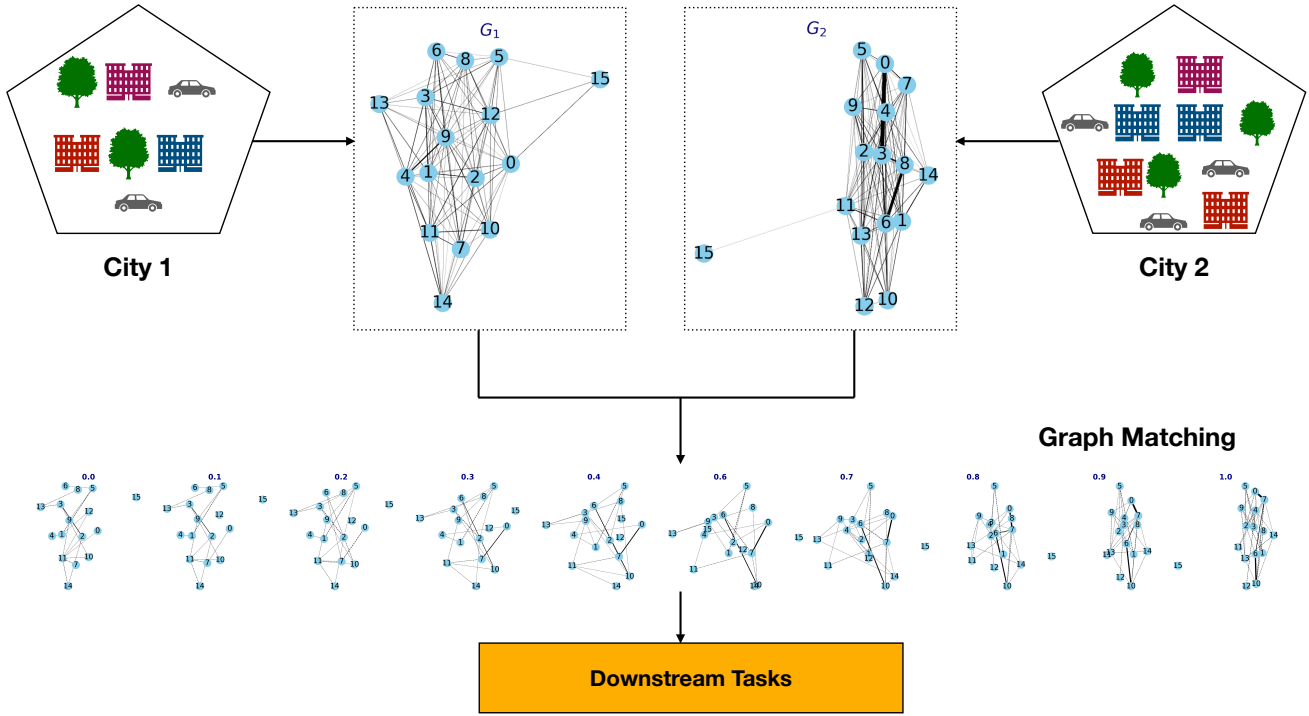

Figure 1: An overview of our method highlighting the graph matching component in the quotient space with mobility networks ($G_1$ and $G_2$) obtained from two cities. The graphs $G_1$ and $G_2$ are matched in the quotient space $\mathcal{G}$. The "walk" from one graph to another is demonstrated here in the graph matching step. The matched graph can then be used for further downstream tasks.

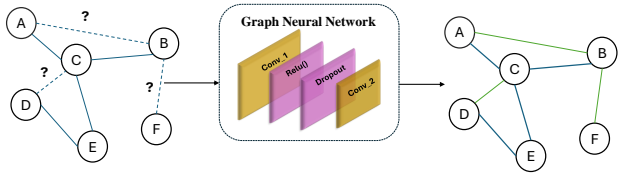

Figure 2: We use GNNs with *conv* layers coupled with *relu* and *dropout* to predict missing links (dashed lines) in the annotated graph of interest on the left. If the link prediction is successful, the model output is the complete graph on the right with the correctly predicted links (in green).

### 2.3 Mobility Prediction

The results of the previous section gives us $G_1$ and $G_2$ - annotated graphs with nodes and edges which are ideal for Graph Neural Networks (GNNs) [19]. GNNs provide an easy way to perform node-level, edge-level, and graph-level prediction tasks while being permutation invariant for structured network data in non-euclidean subspaces. A GNN is an optimizable transform that operate on all attributes of a graph (nodes, edges, global context, inherent relationships etc.) while preserving symmetries and the underlying structure of the data.

The GNN considered is constructed using the "message-passing neural network" (MPNN) backbone [6]. MPNNs follow an iterative scheme to update the nodes by aggregating information from other nearby nodes. The aggregation function is permutation invariant and passes on the aggregated information to the next layer. This type of information flow makes GNNs particularly adept at node categorization as well as link (edge) prediction. For this paper, we are focused on the latter. Given a graph $\{G_\gamma = (E_\gamma, V_\gamma) | \gamma \in \Gamma)\}$, with total $n$ nodes or vertices $V$, and a set of edges $E$, the forward pass through the GNN has two phases - a message passing phase and a readout phase. In the message passing phase for a particular vertex $v \in V$, the message function $M_t$ aggregates the information from its neighbors and updates $v$ according to some update function $U_t$ at any time $t$ as follows:

$$m_i^{t+1} = \sum_{j \in N(i)} M_t \left( h_i^t, h_j^t, e_{ij} \right), \quad (4)$$

$$h_i^{t+1} = U_t \left( h_i^t, m_i^{t+1} \right), \quad (5)$$

where $N(i)$ denotes the neighbors of the $i^{th}$ node in the graph $G_\gamma$, $h$ denotes the hidden state at a particular node, and $e_{ij}$ denotes the edge weight between two nodes $i, j$. The readout phase computes a feature vector for the whole graph using some readout function $R_t$. $M_t, U_t$ and $R_t$ are all learnable differentiable functions that can be estimated during GNN training. We use GNNs to predict missing links (edge) between nodes or vertex of interest as seen in Figure 2.

To study the performance of GNNs on matched graphs we construct a hypothetical where $G_1$ (corresponds to City 1) and $G_2$ (corresponding to City 2) represent two urban cities separated by a span of a few years as mentioned previously. We assume again that $G_1$ is the growing city (in current time) whereas $G_2$ is the city (from the past) that best emulates the current growth of $G_1$. Then, the matched graph $G_{1p}$ can be used to study the behaviour of human mobility and these results can be extrapolated for $G_1$. Specifically missing link prediction helps us understand how the graph will change should new edges be introduced to it.

By splitting $G_{1p}$ (the matched graph) randomly we create a training set (containing "positive" edges) and testing set. We generate random "negative" edges during training which are input to the GNN alongwith the positive edges from the training set. The GNN outputs a label which denotes whether an edge is positive or negative. We use the binary cross-entropy loss function below to train the model:

$$\mathcal{L} = -\sum_{u\sim v\in \mathcal{D}} y_{u\sim v}\log(\hat{y}_{u\sim v}) + (1-y_{u\sim v})\log(1-\hat{y}_{u\sim v}) \quad (6)$$

where $y$ are the ground truth labels, $\hat{y}$ are the predictions made from the GNN and $\mathcal{D}$ denotes the set of all edges positive and negative. This loss function assigns a high score to the positive edges and low scores to negative edges, thus enabling the GNN to identify missing edges.

**Architecture details.** We use the GraphSAGE network [9] for the GNN and train it with an Adam Optimizer [13]i wth a learning rate = 0.01 for 5000 epochs. GraphSAGE leverages node features to learn an embedding function that generalizes to unseen nodes. By incorporating node features in the learning algorithm, it simultaneously learn the topological structure of each node's neighborhood as well as the distribution of node features in the neighborhood. The network consists of *SAGEConv* layers coupled with *relu* and *dropout*as shown in Figure 2 . The depth of the network is adjusted as the total number of nodes increases.

In the following sections we report the graph matching performance and the prediction results using matched vs unmatched graphs.

## 3 Experiment and Results
### 3.1 Dataset
The New York Yellow Taxi Trip Data [1] includes pick-up and drop-off dates/times, pick-up and drop-off locations, trip distances, itemized fares, rate types, payment types, and driver-reported passenger counts. The dataset consists of 112,234,626 trips between 264 unique pick-up and drop-off locations in New York City for yellow taxicabs. For all experiments, $G_1$ and $G_2$ are constructed using the pick-up and drop-off locations as a "node". We also consider two mobility modalities - average travel time and total number of trips between a pair of nodes where these quantities constitute an "edge" between two nodes. The entire dataset is subset to randomly select N unique nodes common between both AM and PM trips. For calculating the value of the edge between a node pair $(n_1, n_2)$ we consider the following mechanisms : (1) for average travel time (in minutes) across trips we sum the total time across all trips and divide by the total number of trips, and (2) for total trips we sum all the trips.

Since this dataset does not contain the geographic coordinates of each drop-off and pick-up location, we use built-in node position method in the `networkx` library using the Fruchterman-Reingold force-directed algorithm [14]. In the following, we choose a subset of the nodes (N = 16) and the average travel time modality for illustration in Figure 3 and Figure 4.

### 3.2 Graph matching
$G_1$ and $G_2$ are distinguished by the time of travel – AM vs. PM. Results are shown in Figure 3 for N = 16 unique nodes that have both AM and PM trips for the average travel time modality. We are able to control the balance between edge and node attributes using the parameter $\lambda$ (from Equation (1)). When $\lambda = 0$ (first row), the graphs are matched without any node attributes (the planar coordinates of the nodes). In the second and third row, the introduction of node attributes improves the graph-matching performance. In Fig 6, we calculate $d_a$ (from Equation(3)) before and after the graph matching process denoted by $d_0$ and $d$, respectively, to demonstrate clearly that the distance between the two graphs reduces after matching further reinforcing the claim that the graphs are now matched.

### 3.3 Edge Prediction
As mentioned previously, we use the graph-matching output ($G_{1p}$) for a link prediction downstream task. When accounting for the different node scales (N = 16, 32, 64, 128), the lesser number of nodes leads to less training data. To subvert this, we perform Monte-Carlo simulations for 100 trials. In each trial, $G_{1p}$ is partitioned into training and test sets randomly and the GNN is trained for 10000 epochs. We then make predictions on the test set (inference) and report the top 10 likelihood scores for each trial. Thus across all 100 trials combined, there are 1000 likelihood scores. During inference, the GNN assigns a likelihood score to the edges in the test set which denotes the confidence of the model that the edge exists in the graph. We report these results in Figure 7. Since GNNs operate on the neighborhood information of a particular node, we expect that for smaller graphs, the predictive network will suffer from the presence of incorrect predictions due to limited data. However, the reverse is also true - scaling the number of nodes should give the GNN enough neighborhood information to make correct predictions with more confidence. We clearly observe this trend in Figure 7, where for N = 128, ≈ 98% of the correct predictions report a very high confidence. While the confidence of the network to make correct predictions is a good metric, we observe in Figure 7 that for some cases (N = 32, 64), the GNN also makes incorrect predictions with high confidence. Thus, we report the actual count of correct vs incorrect predictions in Figures 8, 9. Additionally, we also report the performance of matched graphs in comparison to its unmatched counterpart by repeating the Monte-Carlo simulations with the unmatched graph as input. We observe that as N ↑, the average edges correctly predicted for the matched graphs show excellent performance.

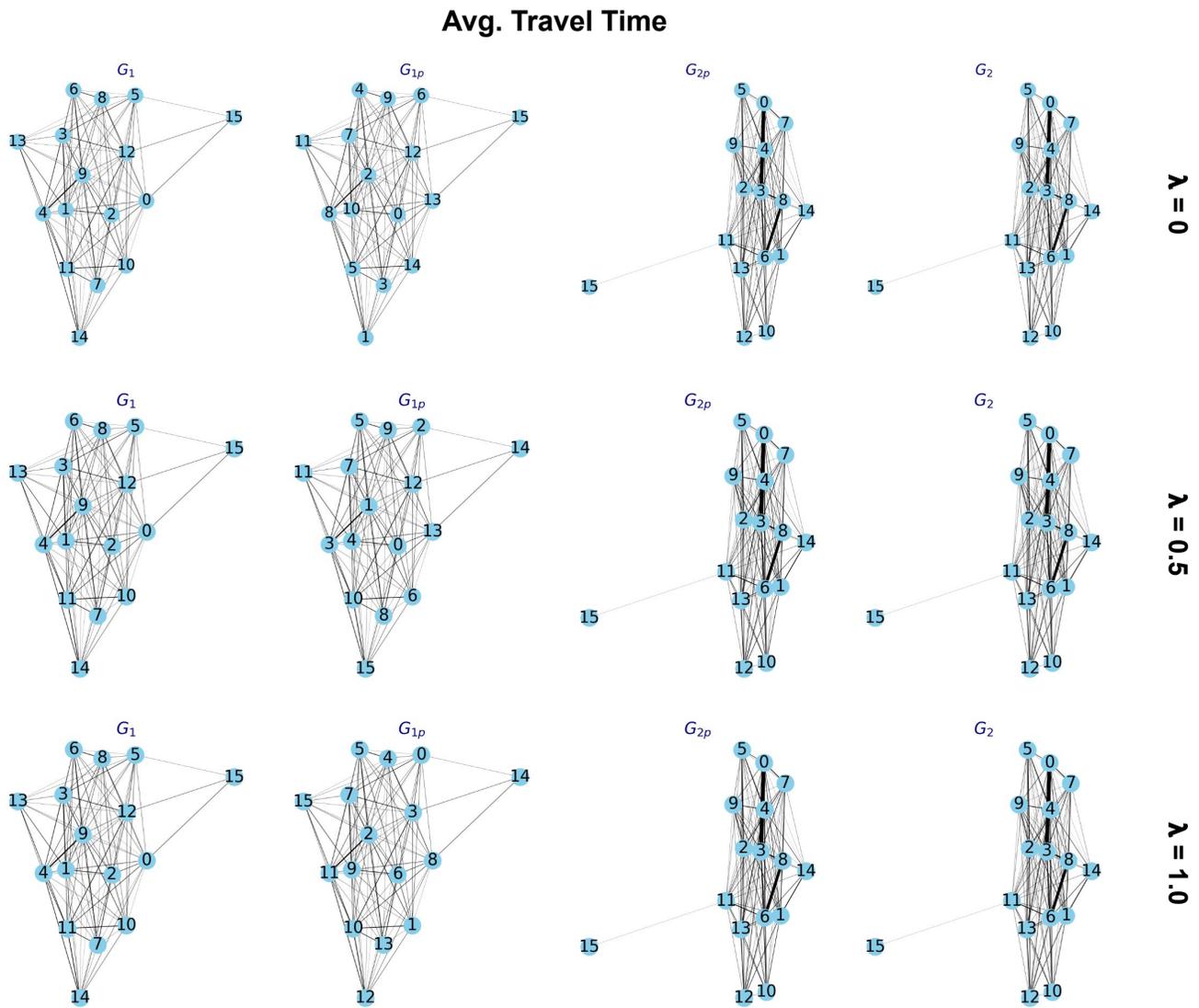

Figure 3: Graph match results for N = 16 unique nodes that have trips in both the AM and PM. $G_1$ and $G_2$ are the original graphs at each end and the matched results are in the middle for each row. The values of $\lambda$ indicate the contribution of node and edge attributes. Since our method affords the flexibility of choice of edge attributes, we chose to represent the average travel time between two nodes as edges. The edges are weighted and a thicker edge denotes a longer travel time between two locations.

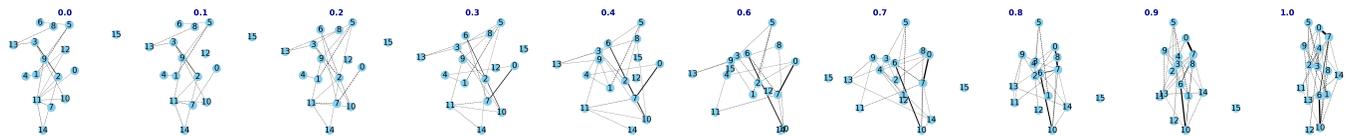

Figure 4: Visual inspection of the interpolation in $\mathcal{G}$ shows that the deformities are more "natural" and a smooth walk between the two original graphs at each end. Dashed lines indicate edges that are changing.

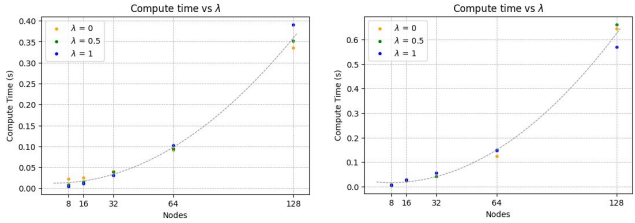

Figure 5: Left: average travel time, right: total number of trips. To analyze the effect of increasing the number of nodes, we visualize the compute time to compare graph-to-graph as a function of number of nodes. Interestingly, we observe that the average compute time follows a quadratic trend (dotted curve) as the number of nodes increases. The fit in (a) and (b) report a r2 score of 0.9988 and 0.999 respectively.

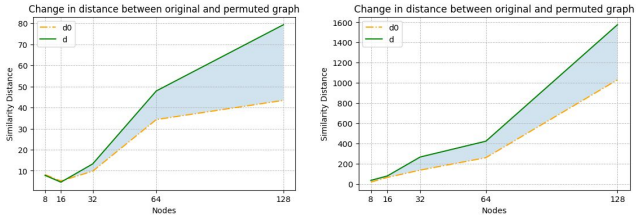

Figure 6: To quantify the similarity distance, we present the average $d$ and $d_0$ values across $\lambda = 0, 0.5, 1$ for the different node sizes across the two modalities (left: average travel time, right: total number of trips). $d$ represents the original distance between the two graphs and $d_0$ is the distance after the matching step. The shaded region represents the change in dissimilarity between the matched and unmatched graphs.

## 4 Conclusion and Future Work

In this work, we explore tools for modeling and visualizing human mobility networks by leveraging the geometric properties of geospatial graphs. We find that the geometric quotient space highlighted in this paper is a robust, scalable, and reliable alternative to traditional graph latent spaces. Additionally, we demonstrate the utility and the compatibility of the quotient space $\mathcal{G}$ with existing downstream machine learning tasks using GNNs. In our experiments we observe that scaling the number of nodes N leads to a significant increase in the performance of the GNN, a crucial finding in the context of real-world mobility networks designed to model human behavior on a large scale in a sustainable manner.

The main drawback of the current model is the lack of real-world coordinate data. We distribute the nodes on the x-y plane using built in methods but integrating geographic coordinates would help translate the results for real-world applications directly. In the future, we aim to explore this possibility by integrating not only geographic coordinates but also socio-economic features at graph nodes (such as population, income class etc.) thus transforming each node feature to be $k$-dimensional (where $k > 1$). In conjunction, we hope to study the ability of the continuous quotient space to provide reliable and unique insights into applications such as link prediction, city planning and expansion, and other downstream tasks.

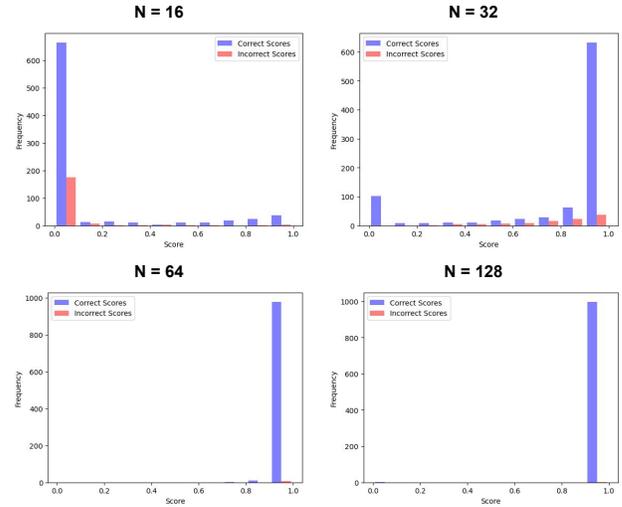

Figure 7: We report the spread of the 1000 likelihood scores by consolidating the top 10 scores from each of the 100 Monte-Carlo trials. For each figure, the x-axis contains the likelihood scores between [0,1] partitioned into 10 bins. The y-axis denotes the total number of incorrect and correct scores that belong to each bin.

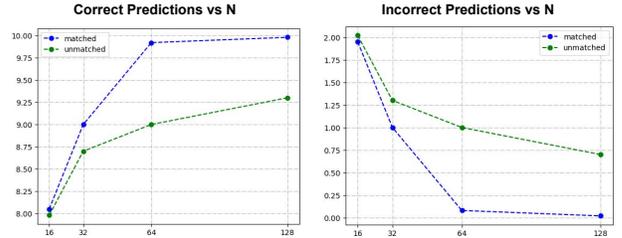

Figure 8: We report the average number of correct (graph on the left) and incorrect (graph on the right) predictions for the matched and unmatched graphs. The x-axis in each figure represents the average number of predictions while the y-axis represents the number of nodes (N). The graphs above are constructed for the GNN trained on the matched graph $G_{1p}$ where the edges constitute the average travel time between two nodes.

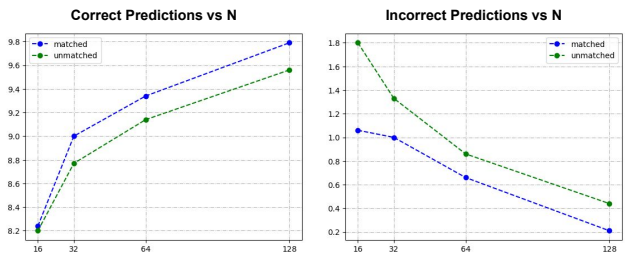

Figure 9: Similar to Figure 8, we also show the predictive performance of the GNN with regards to the matched graph $G_{1p}$ constructed for the total number number of trips modality. In each figure, the x-axis represents the average number of predictions, and the y-axis represents the total number of nodes in $G_{1p}$.